\documentclass[twocolumn]{aastex631}
\usepackage{amsmath}
\usepackage{epsf}
\usepackage{color}
\usepackage{graphicx}
\usepackage{afterpage}
\usepackage{color}
\usepackage{enumitem}
\usepackage{mathtools}
\usepackage[mathcal]{eucal}
\usepackage[utf8]{inputenc}
 \let\mathscr\relax
\usepackage[scr]{rsfso}

\begin{document}

\title{The James Webb Space Telescope Mission}


\author[0000-0003-2098-9568]{Jonathan P. Gardner}
\affiliation{NASA Goddard Space Flight Center, 8800 Greenbelt Rd, Greenbelt, MD 20771, USA}
\email{Email: jonathan.p.gardner@nasa.gov}

\author[0000-0002-6460-0078]{John C. Mather}
\affiliation{NASA Goddard Space Flight Center, 8800 Greenbelt Rd, Greenbelt, MD 20771, USA}

\author{Randy Abbott}
\affiliation{Ball Aerospace \& Technologies Corp., 1600 Commerce Street, Boulder, CO 80301, USA}
\affiliation{Retired}

\author{James S. Abell}
\affiliation{NASA Goddard Space Flight Center, 8800 Greenbelt Rd, Greenbelt, MD 20771, USA}

\author{Mark Abernathy}
\affiliation{Space Telescope Science Institute, 3700 San Martin Drive, Baltimore, MD, 21218, USA}

\author{Faith E. Abney}
\affiliation{Space Telescope Science Institute, 3700 San Martin Drive, Baltimore, MD, 21218, USA}

\author[0000-0001-6559-6616]{John G. Abraham}
\affiliation{NASA Goddard Space Flight Center, 8800 Greenbelt Rd, Greenbelt, MD 20771, USA}

\author[0000-0002-4542-921X]{Roberto Abraham}
\affiliation{Department of Astronomy \& Astrophysics, University of Toronto, 50 St. George Street, Toronto, ON M5S 3H4, Canada}
\affiliation{Dunlap Institute for Astronomy and Astrophysics, University of Toronto, 50 St George Street, Toronto, ON M5S 3H4, Canada}

\author[0000-0001-5625-4091]{Yasin M. Abul-Huda}
\affiliation{Space Telescope Science Institute, 3700 San Martin Drive, Baltimore, MD, 21218, USA}

\author{Scott Acton}
\affiliation{Ball Aerospace \& Technologies Corp., 1600 Commerce Street, Boulder, CO 80301, USA}

% ...

\begin{abstract}
 
 Twenty-six years ago a small committee report, building on earlier studies, expounded a compelling and poetic vision for the future of astronomy, calling for an infrared-optimized space telescope with an aperture of at least $4m$. With the support of their governments in the US, Europe, and Canada, 20,000 people realized that vision as the $6.5m$ James Webb Space Telescope. A generation of astronomers will celebrate their accomplishments for the life of the mission, potentially as long as 20 years, and beyond. This report and the scientific discoveries that follow are extended thank-you notes to the 20,000 team members. The telescope is working perfectly, with much better image quality than expected. In this and accompanying papers, we give a brief history, describe the observatory, outline its objectives and current observing program, and discuss the inventions and people who made it possible. We cite detailed reports on the design and the measured performance on orbit.

\end{abstract}

\section{Introduction}

We summarize the history, concept, scientific program, and technical performance of the James Webb Space Telescope \citep[JWST, Webb;][]{gardner06}. This paper points to and extracts key results from detailed papers in this issue on the four instruments: Near-Infrared Camera \citep[NIRCam;][this issue]{rieke23}, Near-Infrared Spectrograph \citep[NIRSpec;][this issue]{boker23}, Near-Infrared Imager and Slitless Spectrograph \citep[NIRISS][this issue]{doyon23}, and Mid-Infrared Instrument \citep[MIRI;][this issue]{wright23}; the telescope \citep[][this issue]{mcelwain23}; the observatory \citep[][this issue]{menzel23}; the scientific performance \citep[][this issue]{rigby23a}; and the brightness of the sky and stray light \citep[][this issue]{rigby23b}.

\begin{figure}
  \centering
  \includegraphics[width=0.5\textwidth]{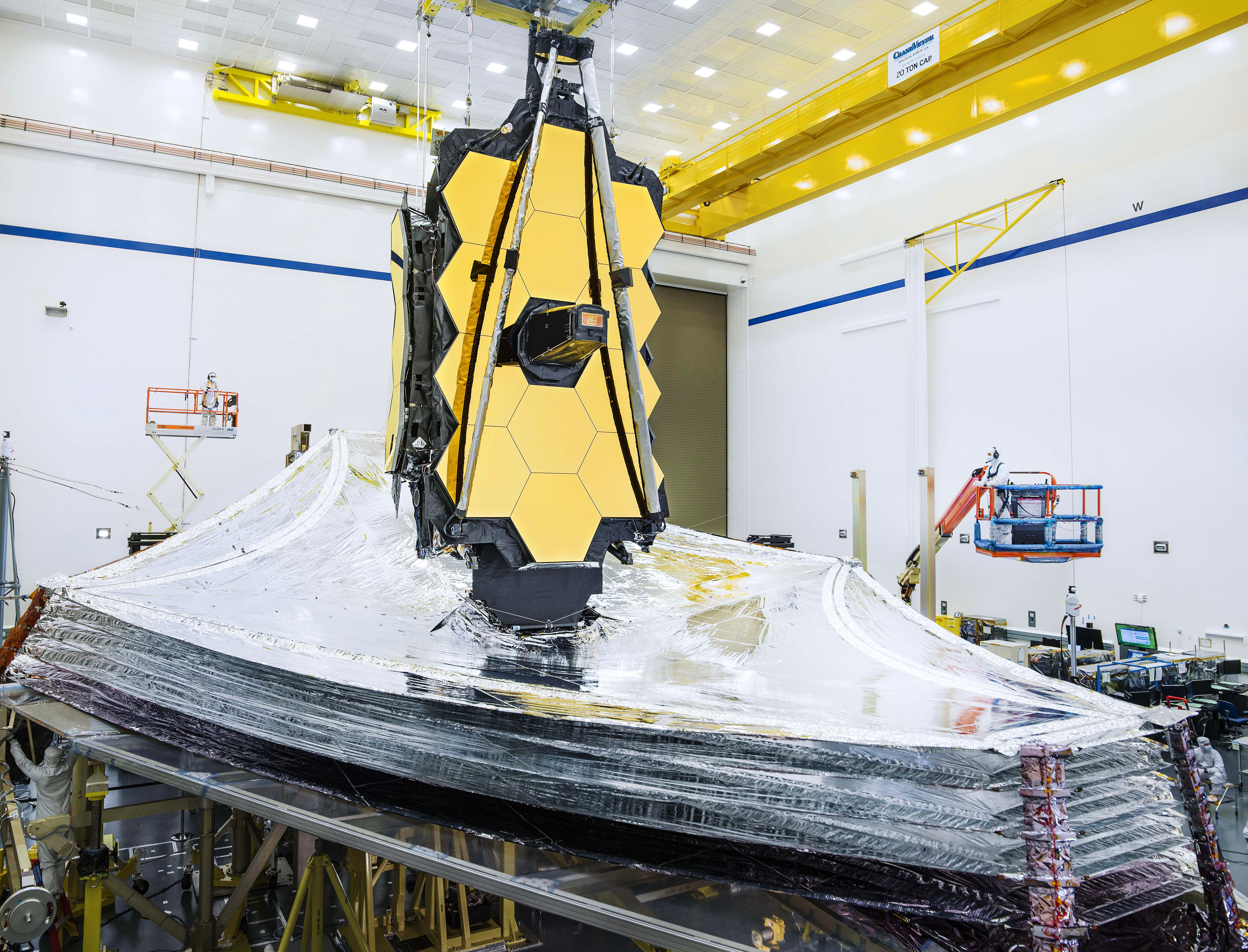}
  \caption{JWST in the Northrop Grumman cleanroom following a sunshield deployment test.}
  \label{fig:sunshield}
\end{figure}

Launched 2021 December 25, the JWST (See figure \ref{fig:sunshield}) is a $6.5m$ diameter cold space telescope with cameras and spectrometers covering $0.6 \mu m$ to $28 \mu m$ wavelength. Orbiting the Sun-Earth L2 point, it extends the discoveries and technologies of the $2.4m$ warm Hubble Space Telescope (HST) and the $85cm$ cold Spitzer Space Telescope. JWST enables observations of the distant early universe, potentially reaching beyond redshift $z \sim 15$, within $\lesssim 300 Myr$ of the Big Bang. JWST’s infrared wavelength range penetrates dust clouds, to observe obscured active galactic nuclei (AGN) and star and planet formation. It shows objects too cool to radiate visible light. It includes the fundamental vibration-rotation bands of important molecules. 

JWST was conceived from the beginning as an international project, led by NASA in partnership with the European and Canadian Space Agencies (ESA and CSA). Observing time allocations are open to all astronomers worldwide. JWST was launched from French Guiana on an ESA-provided Ariane 5 rocket. JWST cost NASA \$8.8B to get to launch, and will make observations for a projected fuel-limited lifetime potentially as long as 20 yrs. JWST’s observations cannot be obtained in any other way: Hubble has 1/6.25 of the collecting area, and, being a room-temperature telescope emits its own infrared beyond $1.7\mu m$. Spitzer was cold but had less than $2\%$ of JWST's collecting area. For ground-based telescopes, the Earth’s warm atmosphere results in a background that is $10^6-10^7$ times higher and also blocks large regions of IR wavelengths.

JWST includes a general-purpose instrument package in its Integrated Science Instrument Module (ISIM). NIRCam covers $0.6\mu m$ to $5\mu m$ with filters, a coronagraph, 0.032 arcsec pixels in its short wavelength channel, and a dichroic filter enabling simultaneous observation in both short and long wavelength bands. NIRSpec covers the same wavelength range with prisms and gratings, with fixed slits, an integral field unit (IFU), and a microshutter array (MSA) to select up to 100 simultaneous targets. NIRISS covers the same wavelength range with slitless spectroscopy, and provides a non-redundant mask for coronagraphy. MIRI covers $5\mu m$ to $28\mu m$ with imaging, IFU spectroscopy, and coronagraphs. The Fine Guidance Sensor (FGS) senses pointing errors and feeds the attitude control system to maintain sharp images. ESA provided the NIRSpec with the detectors and micro-shutter array provided by Goddard, CSA provided the NIRISS and FGS, and a partnership of JPL and a European consortium provided MIRI. NIRCam, and the US portions of MIRI and NIRSpec, were funded by NASA.

JWST’s first year of observations includes nearly half of a year of guaranteed time observations  allocated to the teams building the instruments, 6 interdisciplinary scientists, and the telescope scientist. There is also a set of 13 competitively-selected Director's Discretionary Early Release Science (DD-ERS) programs with data that become public immediately. The remaining observations were selected from over 1200 proposals through a dual-anonymous peer review, in which the identities of the proposers were not known to reviewers. Anyone can propose, regardless of nationality or institution. There were 286 programs selected in Cycle 1, including 263 pointed programs, 3 pure parallel programs, and 20 archival or theory programs. The Cycle 2 call received more than 1600 proposals submitted by more than 5000 scientists from around the world. NASA provides funding for data analysis to US-based investigators who are awarded time on JWST, and for archival and theoretical research related to JWST.

JWST is a technical pioneer, the first deployable segmented space telescope, with mirror segments that are aligned and focused after launch, with its remarkable cryogenic optical system providing diffraction-limited images (Strehl ratio 0.8) as short as $1.1 \mu m$. Its detectors are larger and more sensitive than previous generations. Its 6.0K cryocooler for the mid IR instrument (MIRI) has no expendables and should last the duration of the mission. It uses cryogenic ASICs to control and read the near IR detectors. Its 5-layer sunshield cools the telescope to between 35K and 55K.  Based on JWST’s successes and lessons learned, NASA is positioned to proceed to the 6m near-IR, optical, and UV flagship telescope recommended by the 2020 Decadal Survey \citep{decadal21}, with the ability to directly detect Earth-like planets around Sun-like stars.

\section{Timeline History of JWST}

In this section we present a brief timeline history of the concept, design, construction and testing of JWST.

\subsection{1980s.}

In the 1980s, preliminary work by Garth Illingworth, Pierre Bély, Peter Stockman, and others on an 8m to 10m passively-cooled infrared telescope began at Space Telescope Science Institute (STScI), following Riccardo Giacconi's advice to work on the ``next big thing." Progress was reported at the STScI/NASA Next Generation Space Telescope (NGST) conference, 1989 September 13 to 15 \citep{bely90}. This international conference excited scientists and engineers with the scientific opportunities and technical challenges enabled by a very large cold infrared space telescope.  A decade before that, in 1979, a National Academy of Sciences (NAS) report recommended the Shuttle Infrared Telescope Facility \citep[SIRTF, later Spitzer;][]{decadal83}, even before the IRAS launch (1983 January 25). Goddard built the Cosmic Background Explorer (COBE) and launched it on 1989 November 18 to measure the Big Bang, the cosmic microwave and IR background light, beginning the era of precision cosmology \citep{mather94}. The discovery of the primeval anisotropy led to the standard model of cosmology, in which gravity acting on density fluctuations could produce galaxies. To build the COBE, Goddard also developed expertise with cryogenic instrumentation and test programs that would become essential for Webb.

\subsection{1990s.}

The Hubble was launched 1990 April 24 and its optical error was soon discovered, leading to two breakthroughs: learning to measure the optical error (spherical aberration) with phase diversity, and demonstrating the value of the planned servicing and repair of Hubble using the Space Shuttle in 1993. The development of the phase diversity algorithms was jointly done by STScI, Goddard and JPL \citep{burrows91, fienup93, krist95}. The experience gained by the people working on the servicing missions, instrument upgrades, and spacecraft hardware repairs and replacements would be critically important for the development of the JWST, as many of the younger engineers working on Hubble later became leaders of the Webb team. The Bahcall report \citep[Decadal Survey:][]{bahcall91} endorsed Spitzer (at that time called the Space Infrared Telescope Facility, SIRTF) and declared that the 1990s would be the ``Decade of the Infrared." The AstroTech conference \citep{illingworth91} reviewed possibilities for large telescopes and cited the Bahcall report. In 1993, the Edison mission, a deployable 2m+ diameter IR telescope, was proposed to ESA \citep{thronson96}. Larger format IR detector arrays, in part spurred by the NICMOS development, enabled galaxy surveys \citep[e.g.,][]{gardner93} and other near-IR science with ground-based telescopes.

The Edison concept included passive cooling, and some of the people working on Edison brought that experience to the JWST team later. The Hubble phase retrieval algorithm was published and the Hubble optics were corrected in its first servicing mission. Soon after, NASA commissioned AURA's HST and Beyond committee, chaired by Alan Dressler, to write a report on potential successors to HST \citep{dressler96}. STScI sent a proposal entitled ``High-Z” to NASA, suggesting that a telescope in a 1$\times$3 AU long elliptical orbit would be helpful, to get outside the interference of the zodiacal dust cloud.  In 1995, JPL submitted the MIRORS proposal to NASA for a passively cooled mid IR telescope \citep{wade96}. 

\subsubsection{1995.}

On 1995 October 30, Ed Weiler of NASA HQ left a phone message for John Mather at Goddard, explaining that NASA was starting a study of what was then called the Next Generation Space Telescope (NGST), and inviting his participation. Within the context of the previous studies, and based on a draft of the HST and Beyond report, NGST would be a $4m$ passively-cooled IR telescope making observations at $1 \mu m < \lambda < 5 \mu m$. GSFC chose Bernie Seery to manage the study. The Federal Government shut down 1995 November 14 to 19 and again 1995 December 16 to 1996 January 6. On 1996 January 6 to 8, a record blizzard closed the Washington DC area and nothing moved for a week. As a result Mather did not attend the dramatic AAS meeting in San Antonio. This pivotal start foreshadowed Webb’s future, as its development was to be affected by more storms, government shutdowns, a terrorist attack, an earthquake in Virginia, a lightning strike, a hurricane in Texas, and the COVID pandemic, all of which would occur before Webb was launched.

On 1996 January 15, the Hubble Deep Field image was released at the AAS meeting, completely changing our view of galaxies in the universe \citep{williams96}. On 1996 January 17, Dan Goldin spoke to the AAS, saying, ``Why do you ask for such a modest thing [$4m$]? Why not go after six or seven meters?" He received a standing ovation, which turned out to be Webb’s first (informal) peer review. Dressler also spoke to present the report at the meeting \citep{dressler96}. On 1996 January 22 to 25, Mather attended a conference in Johannesburg, and met ESA counterparts for the first time to discuss ideas for the NGST. The final Dressler report appeared in 1996 May \citep{dressler96}. Wendy Freedman’s Hubble Key Project reported on the Hubble Constant, discussing their initial values (around $60 km s^{-1} Mpc^{-1}$) and comparing to others (up to $80 km s^{-1} Mpc^{-1}$) \citep[][and references therein]{freedman01}. Predictions of early galaxies that could be visible to Webb depend strongly on this number, and would be important when a descope decision was taken in 2001.

\subsubsection{1996 to 1999.}

The year 1996 marked the industrial kickoff. Around May, GSFC presented NGST ideas to interested aerospace companies in a meeting held at the Space Telescope Science Institute (STScI). NASA issued a Cooperative Agreement Notice so the companies could compete for this new work. Two studies were chosen (TRW and Ball) and they and the Government/STScI teams reported in September. Jonathan Gardner, newly hired at Goddard, attended the meeting, learning about NGST for the first time. The reports said they could meet Goldin’s cost target of \$0.5B, though what exactly was included in these estimates is unclear now. Outside this group, hardly anyone believed that such a low cost was possible.

In 1997, NASA solicited proposals for the Ad Hoc Science Working Group to advise NASA on the choice of scientific objectives and instruments; the committee first met later that year. In 1997, AURA issued a report “Visiting a Time When Galaxies were Young”, detailing the concepts that had been developed in 1996 \citep{stockman97}. In 1998, a conference in Liège, Belgium reported progress developing science drivers and technological challenges \citep{liege98}, after ESA established an NGST Task Force. The STScI was officially made the NGST Science Operations Center in a ceremony with Senator Barbara Mikulski and Administrator Goldin. The cosmic acceleration (dark energy) was discovered \citep{riess98, schmidt98, perlmutter99}, adding to the growing sense that NGST was key to studies of the distant universe, as a dark-energy-dominated universe has galaxies forming at higher redshift than in a matter-dominated flat universe. The conference ``Science with the NGST (Next Generation Space Telescope)” was held at GSFC \citep{smith98}.

On July 7, NASA selected Lockheed Martin and TRW to conduct Phase A mission studies, preliminary analysis of the design, and cost. Discussions began with international partners about their possible roles as ESA, CSA and the Japanese Space Agency (JAXA) participated in the ASWG. In September, just as hurricane Floyd approached, the NGST Science and Technology Exposition was held in Hyannis, MA and the choices of possible near IR spectrometer designs were presented \citep{smith00}. \cite{mather00} reported the early history of NGST and listed the instrument studies then underway. The MEMS (micro-electromechanical system) concept for a multi-object spectrometer was favored over a Fourier spectrometer, because the Fourier multiplex advantage does not apply if most of the sky is empty and detector noise is not high \citep{gardner00}. \cite{moseley99} described the microshutter array, which has significant advantages in size, contrast and cryogenic operation over the micromirror arrays developed by Texas Instruments for digital light projection \citep{mackenty00}.

The ASWG final report recommended that NGST include a NIRCam, NIRSpec with a MEMS-based multi-object spectrograph capability and a MIRI doing both imaging and spectroscopy. They gave three options for a fourth instrument: a tunable filter, a visible-light camera and an integral-field spectrograph \citep{stockman01}. By this time, JAXA had decided not to participate in the mission, and the ASWG recommended that NASA partner with ESA and CSA.

\subsection{2000s}

Ed Weiler signed the Formulation Authorization Document in 1999, defining NGST as a NASA priority, but a mid-infrared instrument was a goal rather than a requirement.  The 2000 Decadal Survey ranked NGST \#1 in large space missions \citep{mckee01}. The ranking was predicated on the inclusion of a mid-infrared instrument, which established it as a requirement for the mission. NASA chose an Interim Science Working Group (ISWG) to write the specifications for the instruments and advise about the telescope. NASA and ESA formed a joint Mid-Infrared Steering Committee, which developed a concept for a mid-infrared instrument (MIRI) doing both imaging and spectroscopy. The agencies negotiated a partnership in which ESA would provide the optical bench and NASA would provide the detectors and cooling system.

In 2001, while preparing the statement of work for the solicitation of the main industrial contract, Goddard descoped the telescope from $50m^2$ to $25m^2$ in collecting area, or $8m$ to $6.5m$ diameter. The descope was to address a mismatch between the budget and the scope in the solicitation. In mid-2002, NASA chose TRW to build the observatory, reserving the ISIM to be built by Goddard. NASA renamed NGST for James E. Webb, who was the second administrator of NASA, from 1961 to 1968. Webb was responsible for getting astronauts to the Moon in 8 years, and for expanding NASA’s science program, which built space telescopes, sent probes to Venus and Mars, and started missions to the outer planets \citep{lambright95}.

In 2002, NASA selected a proposal led by Marcia Rieke of the University of Arizona to build the NIRCam. In the same proposal call, NASA selected several members of the flight Science Working Group (SWG; see Table~\ref{tab:swg}), including 6 interdisciplinary scientists and the US members of the MIRI science team. NASA signed MOUs with ESA and CSA detailing the international partnership. ESA would provide the launch vehicle, the NIRSpec and the optical bench for MIRI. CSA would provide the FGS, and install a tunable-filter imager (TFI) on the other side of the FGS optical bench. The development of a cryogenic etalon meeting the requirements for the TFI would prove to be challenging and ultimately was not successful. The SWG also includes NASA Project Scientists and representatives of the partner agencies. 

\begin{deluxetable*}{lll}
\vspace{2mm}
\tabletypesize{\small}
\tablecaption{The JWST Science Working Group}
\tablewidth{\textwidth}
\tablehead{
\colhead{Name} & \colhead{Institution} & \colhead{Position}
}
\startdata
Santiago Arribas&CSIC&NIRSpec Science Representative * \\
Mark Clampin&NASA/HQ&Observatory Project Scientist * \\
René Doyon&Univ of Montreal&CSA Representative \\
Pierre Ferruit&ESA&ESA Representative \\
Kathryn Flanagan&STScI, retired&STScI Representative * \\
Marijn Franx&Leiden University&NIRSpec Science Representative * \\
Jonathan Gardner&NASA/GSFC&Deputy Senior Project Scientist \\
Matthew Greenhouse&NASA/GSFC&ISIM Project Scientist \\
Heidi Hammel&AURA&Interdisciplinary Scientist \\
John Hutchings&DAO, retired&CSA Representative * \\
Peter Jakobsen&ESA, retired&ESA Representative * \\
Jason Kalirai&JHU/APL&STScI Representative * \\
Randy Kimble&NASA/GSFC&Integration, Test, \& Commissioning Project Scientist \\
Nikole Lewis&Cornell University&STScI Representative * \\
Simon Lilly&ETH Zurich&Interdisciplinary Scientist \\
Jonathan Lunine&Cornell University&Interdisciplinary Scientist \\
Roberto Maiolino&University of Cambridge&NIRSpec Science Representative \\
John Mather&NASA/GSFC&Senior Project Scientist \\
Mark McCaughrean&ESA&Interdisciplinary Scientist \\
Michael McElwain&NASA/GSFC&Observatory Project Scientist \\
Matt Mountain&AURA&Telescope Scientist \\
Malcolm Niedner&NASA, retired&Deputy Senior Project Scientist/Technical \\
George Rieke&University of Arizona&MIRI Science Lead \\
Marcia Rieke&University of Arizona&NIRCam Principal Investigator \\
Jane Rigby&NASA/GSFC&Operations Project Scientist \\
Hans-Walter Rix&Max Planck Institute&NIRSpec Science Representative * \\
George Sonneborn&NASA, retired&Operations Project Scientist * \\
Massimo Stiavelli&STScI&Interdisciplinary Scientist \\
H. Peter Stockman&STScI, retired&STScI Representative * \\
Jeff Valenti&STScI&STScI Representative \\
Chris Willott&CNRC&NIRISS Science Lead \\
Rogier Windhorst&Arizona State University&Interdisciplinary Scientist \\
Gillian Wright&UKATC&MIRI European Principal Investigator
\enddata
\tablecomments{The JWST Science Working Group (SWG) consists of GTOs, NASA Project Scientists, and representatives from ESA, CSA and STScI. (*) designates a former member who was replaced by someone else. The NIRSpec Science Representative position rotated through several members of the NIRSpec team. The SWG first met on 2002 August 19 and disbanded at the end of JWST Commissioning, 2022 July 12.}
\label{tab:swg}
\end{deluxetable*}

Phil Sabelhaus was selected as the Project Manager as the project entered its detailed design phase. In late 2003, the part of TRW selected to build JWST was acquired by Northrop Grumman. Northrop, in consultation with Goddard and the optics Product Integrity Team, selected beryllium for the mirror material. The competing glass sandwich technology had failed to meet its figure stability requirements \citep{stahl04}. In addition, the Spitzer Space Telescope (building on other IR space missions) had demonstrated that construction of a high-precision beryllium mirror was possible. In 2003, the fabrication of the flight mirrors began by Northrop through a sub-contract to Ball Aerospace.

Beginning the mirrors was a key step and reflected the expectation that they could be a critical path pacing item. The Spitzer Space Telescope was launched 2003 August 25. Spitzer rapidly demonstrated the power of mid-IR space missions \citep[e.g.,][]{werner04}, and confirmed the wisdom of including the MIRI in JWST. On 2004 March 3, construction officially began for the JWST. In 2005, in order to conserve mass, a helium pulse tube cryocooler to be built by Northrop Grumman was chosen to replace the planned solid hydrogen cooler for the MIRI. \cite{gardner06} published a special issue of the journal Space Science Reviews, detailing the JWST science objectives and design. In 2005 Mike Griffin replaced Sean O’Keefe as NASA Administrator, and in 2007, Griffin approved the contribution of the European Ariane 5 rocket for JWST. The final servicing mission for HST was conducted in 2009, freeing the large cleanroom at Goddard for the integration of the JWST instruments into the ISIM, the integration of the Optical Telescope Element (OTE), and the assembly of the ISIM and the OTE.

\subsection{2010s.}

In 2010, CSA realized that they would not be able to deliver the tunable filter imager on schedule, removed the etalon, and redesigned the camera as the NIRISS \citep{doyon12}. The near-infrared detectors were found to be degrading, and it became clear that they would have to be replaced \citep{rauscher14, rauscher14b}. The spacecraft and sunshield were also behind schedule.

Budget troubles mounted, and Senator Mikulski asked for an independent review and a realistic budget estimate that would not grow every year. Congress was very skeptical of the planned 2014 launch date and the budget. The HgCdTe detectors were found to be degrading in storage, and a new version was ordered. NASA responded to the growing concerns around schedule and cost, by setting up a Test Assessment Team for advice on the testing path forward. In response to Senator Mikulski's request for a full review of what was going wrong, the NASA Administrator set up the Independent Comprehensive Review Panel (ICRP). The ICRP report was hard hitting and direct, confirming the basic soundness of the mission and concurring with the recommendations of Project management that additional budget and schedule were needed. The formal recommendations were all accepted by the Administrator.  The ICRP emphasized the scientific value of JWST, but outlined the budgetary issues that had plagued the Project and recommended that a thorough Joint Cost and Schedule assessment be done quickly.  Project Manager Phil Sabelhaus was replaced by Bill Ochs. In 2011, that budget assessment was finished, and indicated that it would take a total budget of \$8B to complete JWST with a launch in 2018, 4 years later than planned.

Frustration with JWST led to a proposed zero budget by the House Appropriation Sub-Committee on 2011 July 7. A grassroots public effort was undertaken to support Senator Mikulski's efforts to restore the budget. The public support was remarkable. The effort paid off and agreement was reached to restore JWST. JWST's budget of \$8.0B to launch was approved by Congress late in 2011, but with strong language capping the cost. In 2012, the MIRI and NIRISS/FGS instruments arrived at GSFC. A Derecho storm cut off all three electrical power lines to GSFC in the middle of a cryo-vacuum test, but there was no damage to the JWST hardware. In 2013 the NIRCam and NIRSpec arrived at GSFC, and the four instruments were assembled into the ISIM by 2014. Altogether three cryo-vac tests were done on the instrument module \citep{kimble16, greenhouse19}, and it was assembled to the telescope and given vibration and acoustic tests at GSFC. Challenges during the three cryo-vac tests included a major snowstorm, a lightning strike on the building and a fire in the clean room prompting evacuation of the personnel. No people or flight hardware were harmed in these events, which took place between 2014 and 2016.

The eighteen primary mirror segments, the secondary mirror, the tertiary mirror, and the fine-steering mirror are all made of beryllium with a gold optical coating. Of the 18 primary mirror segments, there are 6 each of 3 optical prescriptions, to enable the six-fold symmetry of the primary mirror. Three spare mirrors, one of each optical prescription, were also prepared. In order to meet the $25nm$ root-mean-squared surface figure requirement for each mirror segment, the mirrors were initially polished to a surface accuracy of about $100nm$, using an iterative process. The segments’ surface figures were then measured at the cyrogenic operating temperature in the X-Ray and Cryogenic Facility (XRCF) at NASA’s Marshall Space Flight Center. Using those cryogenic measurements, the segments were polished again so that they would meet the surface figure at their operating temperature. The secondary, tertiary and FSM were treated in a similar manner. The mirrors were then coated, underwent environmental testing, and acceptance testing at the operating temperatures was conducted again in the XRCF. The individual mirrors were complete by early 2012 \citep{feinberg12}.

With the mirrors complete, the hexapod actuators were assembled onto each mirror. Assembly of the primary mirror segments onto the backplane took place at Goddard during late 2015 and early 2016. The integration of the ISIM onto the back of the primary mirror was complete by 2016 May 24, at which time, the assembled Optical Telescope element and ISIM became known as OTIS. Vibration and acoustic testing took place over the following year.

\begin{figure}
  \centering
  \includegraphics[width=0.5\textwidth]{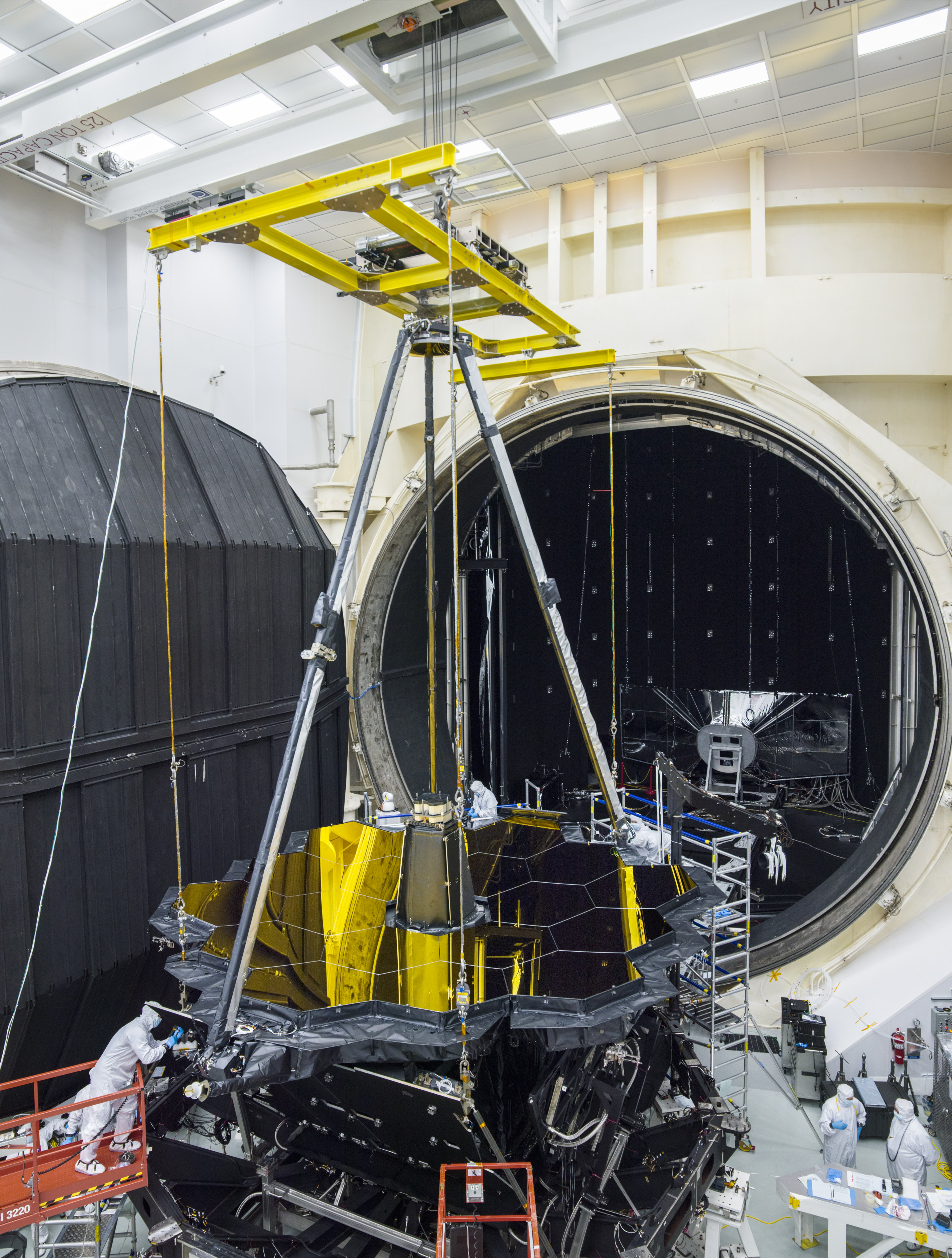}
  \caption{The JWST telescope and ISIM in preparation for the OTIS thermal vacuum test at the Johnson Space Center's historic Chamber A.}
  \label{fig:chambera}
\end{figure}

On 2017 May 7 the OTIS was flown to NASA’s Johnson Space Center on a C5C aircraft, and spent 6 months there, including a 100-day full scale cryo-vac test (see Figure \ref{fig:chambera}), right through Hurricane Harvey and $1.3m$ of rain \citep{kimble18}. On 2018 February 2 the OTIS arrived at Northrop Grumman in Redondo Beach for integration with the warm spacecraft and sunshield, which themselves were undergoing their element-level acoustic and vibration tests \citep[][this issue]{menzel23}. In 2017, delays in the spacecraft (e.g., propulsion system welding problems) and the sunshield development at Northrop began to indicate that a 2018 launch was not likely. By early 2018 it was clear that added time and money was needed, and another Independent Review Team was set up. They emphasized ``mission success" but noted that a launch delay was inevitable for the integration and testing (I\&T) that remained. In 2018 May, loose screws and washers were found after a vibration test of the sunshield, making it clear that a significant delay was necessary to get JWST back on track. Rework of the sunshield followed.  A new launch date in 2020 was selected as likely.

During this decade the Mission Operations Center was being set up at STScI, and a major software development effort was being undertaken by STScI with GSFC management for mission operations, science support and science operations. The JWST Science Advisory Committee (JSTAC) was chartered to help STScI and NASA develop science policies and approaches with the explicit goal of ``maximizing the scientific return" from JWST. The JSTAC met for 8 years in parallel to the SWG. Both advisory committees were eventually succeeded by the JWST Users Committee (JSTUC).

\subsection{2020s.}

Progress was good but the complexities of the I\&T activities led to a further delay into 2021, compounded by the global coronavirus pandemic.  In 2020 March, COVID-19 forced nationwide closures and NASA telework, but aerospace workers at the Northrop Grumman facility in California were permitted to work under enhanced COVID-19 safety procedures. After the final vibration and acoustic test in 2020, the flight transponders failed in early 2021 and were returned to the manufacturer for repair. Flight rehearsals began in the Mission Operations Center in 2020, using the digital twin, an observatory simulator that is still in use today for verification of command sequences. Some of the rehearsals included remote or hybrid participation due to the pandemic; this would prove valuable experience during commissioning.

\begin{figure*}
  \centering
  \includegraphics[width=\textwidth]{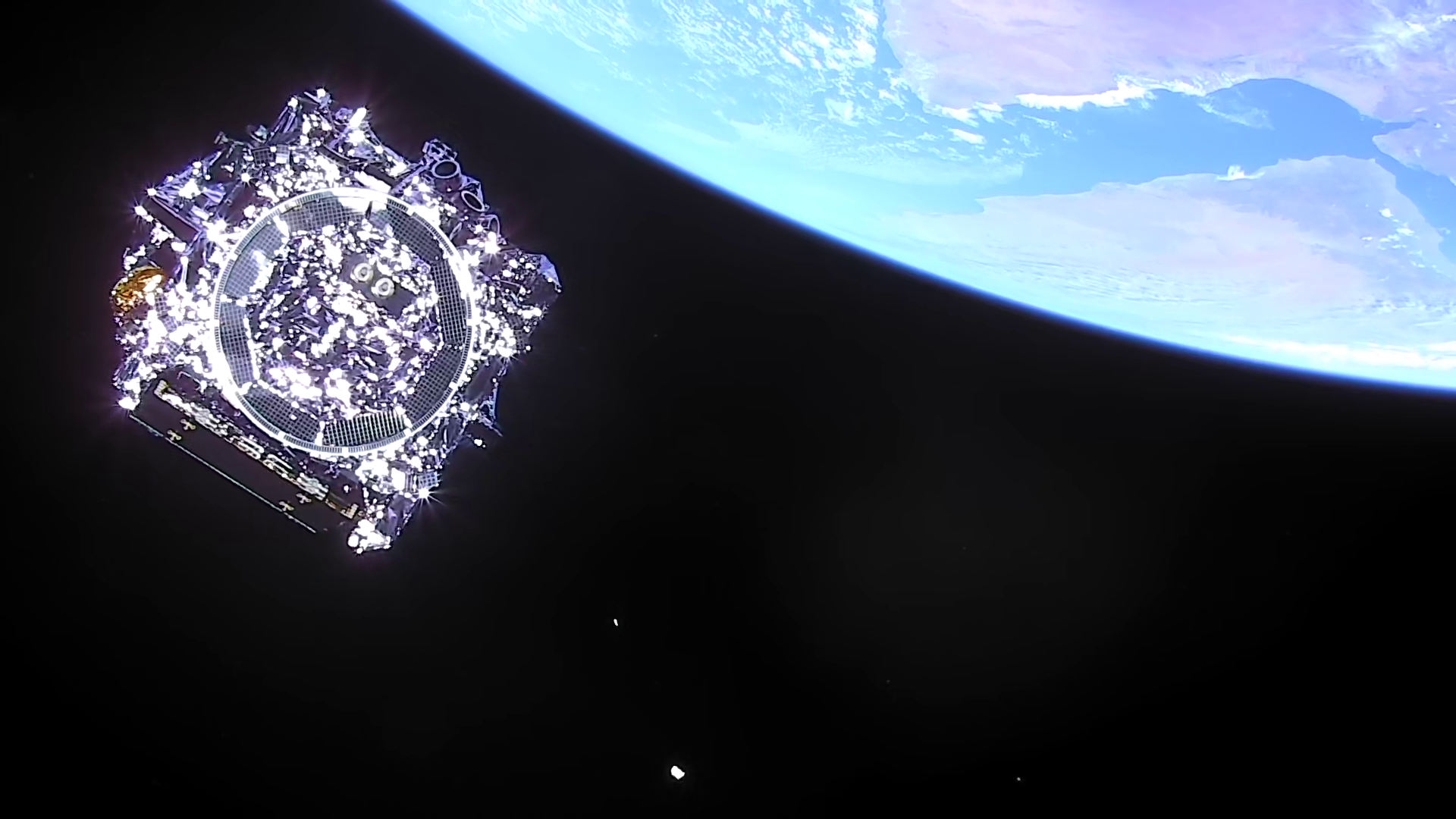}
  \caption{JWST viewed from a camera in the upper stage of the Ariane 5 launch vehicle, just after separation. The Gulf of Aden and the east coast of Africa are visible in the upper right. This still from a video is courtesy ESA and Arianespace. The full video is available at: \url{https://www.esa.int/ESA_Multimedia/Videos/2021/12/Webb_separation_from_Ariane_5}.}
  \label{fig:separation}
\end{figure*}

The final I\&T work in 2021 went extremely well, adding to confidence that the spacecraft and sunshield were becoming mature and ready for launch. On 2021 October 21, JWST arrived at the launch site in French Guiana on the MN Colibri ship, after passing through the Panama Canal. On 2021 December 25 at 12:20 UTC, the Ariane 5 launched the Webb exactly as planned, with a flawless launch and positioning of JWST for its trajectory to L2 (see figure \ref{fig:separation}). The launch went so well that the propellant needed to adjust its trajectory to L2 and the insertion were much less than budgeted. The propellant available for orbit adjustment will likely allow a mission life extension to a predicted 20 years, far larger than the 10-year goal. 

\subsubsection{2022.}

The first two weeks of ``deployment terror" went remarkably well, with almost no unexpected issues. In particular, the sunshield with its 140 release mechanisms deployed successfully. The years of careful testing and checking had paid off. The following two weeks of mirror deployments also went smoothly and JWST was inserted into its L2 orbit 29 days after launch, ready for the slow, crucial process of aligning the mirrors. The outcome was that the optical performance and stability exceeded the requirements, with $1.1 \mu$m diffraction-limited imaging at NIRCam, almost twice as good as the requirement ($2 \mu$m) \citep[][this issue]{mcelwain23}. 

Despite the spike in COVID-19 cases at the end of 2021, the careful protocols developed and rehearsed by the Mission Operations Team enabled staffing of all critical ground-support positions throughout launch and early commissioning. Precautions included an expansion of the space used by the Mission Operations Center (MOC) at STScI, to permit socially-distanced staffing; greater reliance on remote support; mandated vaccination or testing, mandated masking, and electronic contact tracing for personnel in the MOC; and rapid antigen testing every other day for on-site staff. 

\begin{figure}
  \centering
  \includegraphics[width=0.5\textwidth]{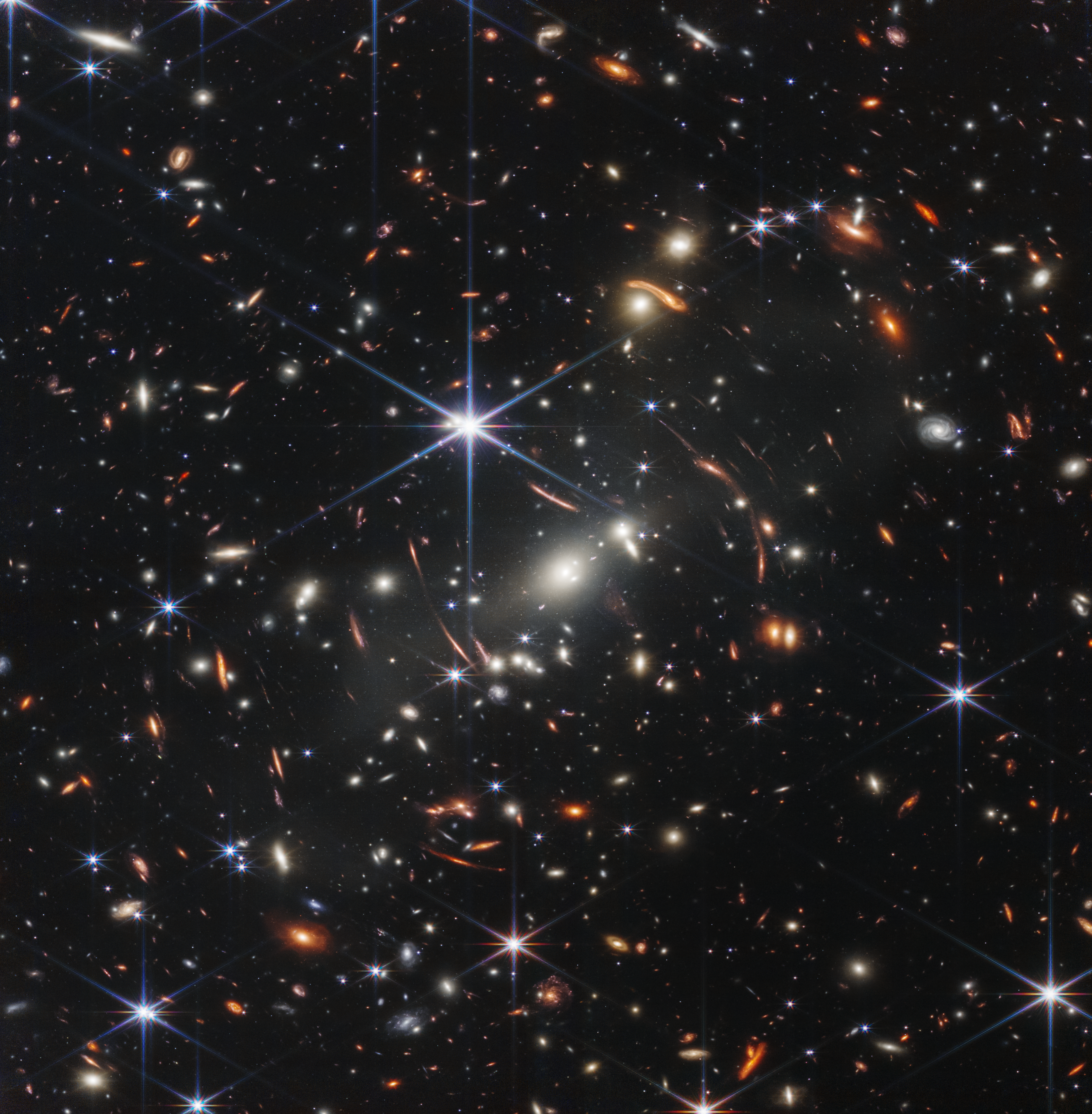}
  \caption{JWST's first released image, the lensing cluster SMACS 0723, was the deepest infrared image ever taken.}
  \label{fig:smacs}
\end{figure}

Overall, commissioning the observatory went extremely smoothly in the MOC, due to the prior tests and rehearsals, and especially the competence, focus and leadership throughout the NASA, Northrop, Ball, STScI, instrument teams, and other contractor teams. Everything on JWST works! Commissioning was completed and all 17 instrument modes were approved for scientific use by 2022 July 10. During commissioning, 120 hours were devoted to Early Release Observations \citep[ERO][]{pontoppidan22}. The ERO of SMACS 0723 was announced by President Biden at the White House July 11 (Figure \ref{fig:smacs}), and Stephan's Quintet, the Carina Nebula, the Southern Ring Nebula, and WASP-96b were released on July 12. The data were released through the archives at the Mikulski Archive for Space Telescopes (MAST) on 2022 July 14; the release also included the commissioning data and online documentation of the scientific performance.

\section{Science Scope}


JWST was designed to address four science themes \citep{gardner06}, which trace cosmic history from the Big Bang to planets conducive to life. The End of the Dark Ages: First Light and Reionization theme seeks to identify the first luminous sources to form and to determine the ionization history of the early universe. The Assembly of Galaxies theme seeks to determine how galaxies and the dark matter, gas, stars, metals, morphological structures, and active nuclei within them evolved from the epoch of reionization to the present day. The Birth of Stars and Protoplanetary Systems theme seeks to unravel the birth and early evolution of stars, from infall on to dust-enshrouded protostars to the genesis of planetary systems. The Planetary Systems and the Origins of Life theme seeks to determine the physical and chemical properties of planetary systems including our own, and investigate the potential for the origins of life in those systems.

Observing time on JWST has been allocated by three methods: Guaranteed Time Observations (GTO), General Observers (GO) time, and Director’s Discretionary (DD) time. The GTO time consists of 4020 hours of JWST observations over the first three years, allocated to the four instrument teams and other scientists on the Science Working Group. The DD time consists of 10\% of the time available for the lifetime of the mission. The GO time is the remainder.

The GTO allocations are listed in Table~\ref{tab:gto}. Changes from the original allocations are due to programs shared between the GTOs and to changes in the overheads that occurred after the selected observations were specified prior to the Cycle 1 Call For Proposals. The GTO scientists chose to allocate the majority of the GTO time in Cycle 1, leaving just 196.1 hours of observation time for Cycles 2 and 3. The table shows the amount of time executed as of 2022 September 30. Further updates are available from https://www.stsci.edu/cgi-bin/get-jwst-gto-time.

\begin{deluxetable*}{llrrrr}
\vspace{2mm}
\tabletypesize{\small}
\tablecaption{GTO Allocations}
\tablewidth{\textwidth}
\tablehead{
\colhead{Team} & \colhead{PI} & \colhead{Original} & \colhead{Current} & \colhead{Remaining} & \colhead{Executed}
}
\startdata
MIRI STScI & Christine Chen & 12.0 & 10.5 & 1.5 & 0.0 \\
NIRISS Team & Rene Doyon & 450.0 & 469.2 & 6.9 & 34.9 \\
NIRSpec Team & Pierre Ferruit & 900.0 & 829.8 & 59.4 & 57.2 \\
MIRI STScI & Scott Friedman & 12.0 & 15.5 & 0.9 & 0.0 \\
MIRI STScI & Karl Gordon & 12.0 & 0.0 & 0.0 & 0.0 \\
MIRI US & Thomas Greene & 60.0 & 75.0 & 0.2 & 15.8 \\
IDS & Heidi Hammel & 110.0 & 141.5 & 1.0 & 32.7 \\
MIRI STScI & Dean Hines & 12.0 & 15.8 & 0.1 & 1.3 \\
IDS & Simon Lilly & 110.0 & 115.9 & 0.1 & 18.6 \\
IDS & Jonathan Lunine & 110.0 & 110.5 & 2.0 & 39.4 \\
IDS & Mark McCaughrean & 110.0 & 100.4 & 33.3 & 12.0 \\
MIRI US & Margaret Meixner & 60.0 & 76.2 & 0.3 & 26.6 \\
Telescope Scientist & Matt Mountain & 210.0 & 212.9 & 10.4 & 9.5 \\
MIRI STScI & Alberto Noriega-Crespo & 12.0 & 10.7 & 0.1 & 0.0 \\
MIRI US & Michael Ressler & 60.0 & 63.5 & 0.7 & 2.4 \\
MIRI Science Lead & George Rieke & 210.0 & 183.1 & 39.2 & 0.0 \\
NIRCam Team & Marcia Rieke & 900.0 & 1041.2 & -3.9 & 108.3 \\
IDS & Massimo Stiavelli & 110.0 & 110.0 & 25.7 & 5.1 \\
IDS & Rogier Windhorst & 110.0 & 122.0 & 0.2 & 23.4 \\
MIRI Europe & Gillian Wright & 450.0 & 452.0 & 18.0 & 62.9 \\
Totals &  & 4020.0 & 4155.7 & 196.1 & 449.8
\enddata
\tablecomments{The time allocations to GTO teams in hours. IDS stands for Interdisciplinary Scientist on the Science Working Group. The original allocations to each team (as set by policy and international agreements) have been modified due to exchanges or collaborations between GTO teams, and due to changes in overhead calculations since the original programs were submitted. The column labeled “Remaining” refers to the number of hours remaining to be allocated in Cycle 2 or Cycle 3. The number of hours in the final column are those that have been executed by 2022 September 30. The current values of the information listed in this table are available from \url{https://www.stsci.edu/cgi-bin/get-jwst-gto-time}}
\label{tab:gto}
\end{deluxetable*}

The DD time is allocated by the Director of JWST’s Science and Operations Center (SOC), located at the Space Telescope Science Institute. In 2018, following a recommendation by the JSTAC, Director Ken Sembach allocated up to 525 hours of Cycle 1 DD time to the DD-Early Release Science Program \citep{levenson18}. The DD-ERS program was designed and implemented by Janice Lee, Jennifer Lotz and Neill Reid. It consists of 13 peer-review selected programs designed to demonstrate JWST’s capabilities and provide open-access data to the community early in the mission. The DD-ERS programs were intended to be conducted within the first 3 to 5 months of Cycle 1; Table~\ref{tab:dders} lists the programs.

\begin{deluxetable*}{llp{0.45\linewidth}p{0.22\linewidth}p{0.12\linewidth}}
\vspace{2mm}
\tabletypesize{\small}
\tablecaption{DD-ERS Programs}
\tablewidth{\textwidth}
\tablehead{
\colhead{ID} & \colhead{Acronym} & \colhead{Title} & \colhead{PI and Co-PIs} & \colhead{Instruments}
}
\startdata
1288 & PDRs4All & Radiative Feedback from Massive Stars as Traced by Multiband Imaging and Spectroscopic Mosaics & Olivier Berne, Emilie Habart, Els Peeters & M, NC, NS\\
1309 & IceAge & IceAge: Chemical Evolution of Ices during Star Formation & Melissa McClure, Abraham C. Boogert, Harold Linnartz & M, NC, NS\\
1324 & GLASS & Through the Looking GLASS: A JWST Exploration of Galaxy Formation and Evolution from Cosmic Dawn to Present Day & Tommaso Treu  & NC, NI, NS\\
1328 & GOALS-JWST & A JWST Study of the Starburst-AGN Connection in Merging LIRGs & Lee Armus, Aaron Evans & M, NC, NS\\
1334 &  & The Resolved Stellar Populations Early Release Science Program & Daniel Weisz & NC, NI\\
1335 & Q-3D & Q-3D: Imaging Spectroscopy of Quasar Hosts with JWST Analyzed with a Powerful New PSF Decomposition and Spectral Analysis Package & Dominika Wylezalek, Sylvain Veilleux, Nadia Zakamska & M, NS\\
1345 & CEERS & The Cosmic Evolution Early Release Science (CEERS) Survey & Steven Finkelstein & M, NC, NS\\
1349 & WR DustERS & Establishing Extreme Dynamic Range with JWST: Decoding Smoke Signals in the Glare of a Wolf-Rayet Binary & Ryan Lau & M, NI\\
1355 & TEMPLATES & TEMPLATES: Targeting Extremely Magnified Panchromatic Lensed Arcs and Their Extended Star Formation & Jane Rigby, Joaquin Vieira & M, NC, NS\\
1364 &  & Nuclear Dynamics of a Nearby Seyfert with NIRSpec Integral Field Spectroscopy & Misty Bentz & NS\\
1366 &  & The Transiting Exoplanet Community Early Release Science Program & Natalie Batalha, Jacob Bean, Kevin Stevenson & M, NC, NI, NS\\
1373 &  & ERS Observations of the Jovian System as a Demonstration of JWST’s Capabilities for Solar System Science & Imke de Pater, Thierry Fouchet  & M, NC, NI, NS\\
1386 &  & High Contrast Imaging of Exoplanets and Exoplanetary Systems with JWST & Sasha Hinkley, Andrew Skemer, Beth Biller & M, NC, NI, NS\\
\enddata
\tablecomments{The 13 selected Director’s Discretionary – Early Release Science Programs. Instruments: NC = NIRCam, NI = NIRISS, NS = NIRSpec, M = MIRI.}
\label{tab:dders}
\end{deluxetable*}

Observing time on JWST is allocated by wall-clock time; all overheads are included in the time allocated. Overheads that are determined by the observing sequence, such as slews or mechanism movements, are accounted by the planning software. Overheads that are independent of the specific observations, such as the calibration program or station-keeping maneuvers, are accounted by a percentage of time applied to each program.

The Cycle 1 GO program allocated approximately 6000 hours. Between the GTO, DD and GO programs, the total number of hours in Cycle 1 exceeds 1 year (8760 hours) by about 25\%. The field of regard of JWST is about 40\% of the sky at any given time. There will need to be enough targets within the field of regard at the end of Cycle 1 to efficiently schedule the observations, and the extra 25\% tail will ensure a smooth transition from the end of Cycle 1 to the beginning of Cycle 2. The remaining allocated Cycle 1 observations will be conducted during the early months of Cycle 2, and there will be a similar transition from Cycle 2 to Cycle 3.

\subsection{Director’s Discretionary Early Release Observation programs}
The DD-ERS program consists of 13 programs ranging from the early Universe to our Solar System. Collectively the programs use most of the JWST instrument modes; the data taken have no exclusive access period so that prior to writing Cycle 2 observing proposals, the scientific community were able to obtain JWST data from the archive that is relevant to their proposals. In this section we describe the DD-ERS programs, and report on those that had early scientific results that were published or submitted by 2022 September 30.

\subsubsection{DD-ERS 1288; PDRs4All: Radiative Feedback from Massive Stars}
Massive stars produce intense winds and radiation which ionizes and heats the surrounding molecular cloud material, affecting future star formation in the cloud. These interactions create Photo-Dissociation Regions (PDRs), which dominate the infrared spectra of star-forming galaxies and drive the evolution of star formation from interstellar matter. PDRs4All \citep{berne22} consists of NIRCam and MIRI imaging and MIRI MRS and NIRSpec IFU spectroscopy of PDRs in the Orion Bar.

\subsubsection{DD-ERS 1309; IceAge: Chemical Evolution of Ices During Star Formation}

Icy grain mantles hold volatile elements and pre-biotic complex organic molecules in star-forming regions. IceAge \citep{mcclure18, mcclure22} will use NIRSpec and MIRI spectroscopy and NIRCam wide-field slitless spectroscopy to study ice chemistry in a representative low-mass star-forming region at several stages: pre-stellar core, Class 0 protostar, Class I protostar and protoplanetary disk. The program will map the spatial distribution of ices to 20 to 50 AU. If the organic molecules survive protostellar infall intact, then the molecular cloud could provide the precursors of biomolecules to the planetary systems that form within it, which could mean that life is a common outcome of star and planetary system formation. 

\subsubsection{DD-ERS 1324; Through the looking GLASS: a JWST exploration of galaxy formation and
evolution from cosmic dawn to present day}

Gravitational lensing by a Frontier Field cluster allows JWST to reach intrinsically faint galaxies in the epoch of reionization. GLASS \citep{treu22.1} will observe Abell 2744 with NIRSpec multi-object spectroscopy and NIRISS wide-field slitless spectroscopy, and will also create a parallel deep field with NIRCam imaging. The parallel field will contain the deepest extragalactic data of the DD-ERS program; the spectroscopy will also be the deepest DD-ERS data obtained in those modes.

GLASS will identify galaxies within the epoch of reionization and measure Ly$\alpha$ velocities and morphologies, rest-frame UV and optical emission line fluxes and UV/optical photometry and sizes.  The program will map metallicity, dust and star-formation rate for galaxies spanning log $M_{\star} \sim 6$ to $10$ at $z>2$, when disks and bulges emerged and feedback was most active. The GLASS NIRISS observations will map systems at $z<3.5$ and log $M_{\star} \gtrsim 6$. The higher resolution NIRSpec observations will spectrally resolve key diagnostic lines which are blended at the lower resolution of the grism spectra. NIRSpec will also reach beyond $z \gtrsim 4$ galaxies.

The NIRCam parallel observations will cover $\sim 18 arcmin^2$ in two regions. They will use 7 broadband filters from F090W to F444W, and will reach 5$\sigma$ limits of 29.2 to 29.7 AB magnitudes (according to the pre-launch exposure time calculator.)

The GLASS team has published a series of papers with initial results, and other groups have also used these public data. The GLASS team papers include: (1) lensed galaxies at $z>7$ \citep{glass01}, (2) photometry and catalogs \citep{glass02}, (3) galaxies at $z>9$ \citep{glass03}, (4) metallicity in low-mass galaxies at $z \sim 3$ \citep{glass04}, (5) the size-luminosity relation at $z>7$ \citep{glass05}, (6) measurements of rest-frame optical lines \citep{glass06}, (7) globular clusters at $z \sim 4$ \citep{glass07}, (8) the detection of a lensed star at $z=2.65$ \citep{glass08}, (9) spectra of low-mass galaxies at $z \gtrsim 2$ \citep{glass09}, (10) measurements of the rest-frame UV at $7<z<9$ \citep{glass10}, (11) masses and M/L at $z>7$ \citep{glass11}, (12) the morphology of high-z galaxies \citep{glass12}, (13) faint cold brown dwarfs \citep{glass13}, (14) a morphological atlas at $1<z<5$ \citep{glass14}, (15) faint high-z sources are intrinsically blue \citep{glass15}, (16) UV slopes at $4<z<7$ \citep{glass16}, and (17) star-formation histories at $5<z<7$ \citep{glass17}.

Other results from the GLASS observations include a lensing model of Abell 2722 \citep{bergamini22} and two studies using ALMA data of $z>12$ candidates \citep{bakx22, popping22}.

\subsubsection{DD-ERS 1328; A JWST Study of the Starburst-AGN Connection in Merging LIRGs}

Luminous Infrared Galaxies (LIRGs) are some of the most active regions of star formation in the Universe. The connection between starbusts and Active Galactic Nuclei (AGN), and the role of galaxy merging in feeding both the star formation and supermassive black hole growth, has implications for both feedback and the star-formation history of the Universe. Program 1328 \citep[e.g.,][]{lai22} will obtain NIRSpec and MIRI IFU spectroscopy, along with NIRCam and MIRI imaging, of active galaxies taken from the Great Observatories All-Sky LIRG Survey \citep[GOALS]{armus09}.

Initial results from GOALS-JWST include detection of an AGN-driven outflow from the nucleus of NGC 7469 using mid-IR spectroscopy \citep{armus22}, and a map of the Polycyclic Aromatic Hydrocarbon (PAH), molecular gas emission \citep{lai22}, and ionization states \citep{u22} in the object. NIRCam and MIRI imaging of NGC 7469 show star-forming regions consistent with young ($<5$ Myr) stellar populations, showing an age bimodality in the star-forming regions of the ring \citep{bohn22}. GOALS-JWST observations of the merger VV114 resolved its double nuclei and showed that the south-western core had AGN-like colors while the north-eastern core was a starburst \citep{evans22}. About half of the mid-IR emission in the object was diffuse, including PAH emission. Observations of the merging galaxy IIZw096 showed that between 40 at 70 percent of the IR bolometric luminosity came from a single region smaller than 175 pc in radius \citep{inami22}.

\subsubsection{DD-ERS 1334; The Resolved Stellar Populations Early Release Science Program}

The Resolved Stellar Populations DD-ERS program \citep{gilbert18, weisz23} will observe the globular cluster M92, the ultra-faint dwarf Draco II and the star-forming dwarf WLM to measure the sub-Solar mass stellar initial mass function (IMF), extinction maps, evolved stars, proper motions and globular clusters. The program will use NIRCam imaging with NIRISS imaging in parallel, both using the F090W and F150W filters, with either wide or medium-band filters in the long-wavelength NIRCam channel. The program will also develop point-spread-function-fitting software specific to NIRCam and NIRISS for evaluating crowded stellar populations \citep{warfield23}.

Early results from Program 1334 include color-magnitude diagrams of M92, reaching almost to the bottom of the M92 main sequence ($\sim 0.1 M_{\odot}$), and finding white dwarf candidate members of M92 in the brightest portion of the white dwarf cooling sequence \citep{nardiello22}.

\subsubsection{DD-ERS 1335; Q-3D: Imaging Spectroscopy of Quasar Hosts with JWST Analyzed with a
Powerful New PSF Decomposition and Spectral Analysis Package}

The Q-3D program uses NIRSpec and MIRI IFU observations of luminous quasars as templates to develop a PSF decomposition and spectral analysis packages, separating the bright central quasar from the host galaxy extended emission \citep{wylezalek22}. The program will measure the stellar, gas and dust components to determine the impact of luminous quasars on their hosts. It will observe three systems: F2M1106, XID2028 and SDSSJ1652. 

\subsubsection{DD-ERS 1345; The Cosmic Evolution Early Release Science (CEERS) Survey}

The statistical study of galaxy formation and evolution through deep-field observations with Hubble, Spitzer, Chandra and many other facilities has a rich history, including the Hubble Deep Field \citep{williams96}, the Great Observatories Origins Deep Survey \citep{giavalisco04}, the Hubble Ultra-Deep Field \citep{beckwith06}, The Cosmic Assembly Near-infrared Deep Extragalactic Legacy Survey \citep[CANDELS,][]{grogin11, koekemoer11}, the Frontier Fields \citep{lotz17} and many others. The CEERS survey will image 100 arcmin$^2$ within the CANDELS Extended Groth Strip field with NIRCam and MIRI. The program will obtain NIRSpec MSA and NIRCam WFSS spectroscopy of objects detected in the imaging. The science goals of CEERS includes finding galaxies at $z>9$ and constraining their nature and abundance, obtaining spectra of galaxies at $z>3$, including candidates at $z>6$, and characterizing the MIR emission from galaxies to study dust-obscured star formation and supermassive black hole growth.

Early results from the CEERS data include the discovery of candidate galaxies at $z>8$ \citep[e.g.,][]{finkelstein22, naidu22, topping22, ono22}, studies of galaxies at $3<z<7$, including quiescent galaxies \citep{carnall22}, AGN host galaxies \citep{kocevski22, onoue22, ding22}, star-forming clumps \citep{chen22a}, and sources detected at other wavelengths \citep[e.g.,][]{chen22b}.

\subsubsection{DD-ERS 1349; Establishing Extreme Dynamic Range with JWST: Decoding Smoke Signals
in the Glare of a Wolf-Rayet Binary}

Colliding-wind Wolf-Rayet (WR) binaries efficiently produce dust, and are important sources for the production of dust in galaxy evolution. WR DustERS \citep{lau22} will observe two carbon-rich WR systems; WR-140 is the archetypal colliding-wind binary, while WR-137 is a known periodic dust-maker \citep{lau20}. The program will obtain the first resolved mid-infrared spectrum of the dust around a carbon-rich WR star, using MIRI IFU and imaging. They will develop PSF-subtraction techniques for observing faint extended emission around bright sources in IFU datasets. The observations of WR-137 will be done with NIRISS AMI mode to detect the faint dust spiral around the bright central source.

MIRI imaging of WR-140 detected 17 nested dust shells, which formed at each periastron 7.93 years apart over the past 130 years. Understanding the chemical properties and spectral signatures of dust formed by binaries like WR-140 is important given their potential role as dust sources in the inter-stellar medium. MIRI spectroscopy of the second dust shell confirmed the survival of carbonaceous dust grains seen as PAH features, and are consistent with a composition of carbon-rich aromatic compounds in a hydrogen-poor environment. Since this carbonaceous dust has lasted for at least 130 years in the harsh radiation environment of the central binary system, it could be a possible early and potentially dominant source of organic compounds and dust in the ISM of our galaxy \citep{lau22}.

\subsubsection{DD-ERS 1355 - TEMPLATES: Targeting Extremely Magnified Panchromatic Lensed Arcs
and Their Extended Star formation}

Objects that are magnified by gravitational lensing can be observed with greater intrinsic spatial resolution and higher sensitivity. TEMPLATES \citep{rigby20} will obtain NIRSpec and MIRI IFU spectroscopy and imaging of 4 gravitationally-lensed galaxies selected at $1<z<4$. The program will spatially resolve star formation structures in an extinction-robust manner, mapping H$\alpha$, Pa$\alpha$ and $3.3\mu m$ PAH features within the galaxies. The selected targets are the brightest and best-characterized lensed systems known. 

\subsubsection{DD-ERS 1364 - Nuclear Dynamics of a Nearby Seyfert with NIRSpec Integral Field Spectroscopy}

Measuring the mass of a super-massive black hole in a galaxy can be done with dynamical measurements or with reverberation mapping. It is important to connect dynamical measurements, primarily done on local quiescent galaxies, to reverberation mapping of more distant active galaxies. Program 1364 will use NIRSpec IFU observations of NGC 4151 to directly measure the mass of the central black hole and compare to previous reverberation mapping measurements of the same object \citep{bentz22}. The program will measure kinematic maps of the stars and gas, intensity maps of the gas and stellar dynamical models of the galaxy.

\subsubsection{DD-ERS 1366: The Transiting Exoplanet Community Early Release Science Program}

Transiting exoplanets will allow JWST to measure atmospheric compositions, structures and dynamics in unprecedented detail. Program 1366 will use time-series observations in all four instruments to observe  WASP-39b in transit, a MIR phase curve of WASP-43b, and a secondary eclipse of WASP-18b. 

Early results include the first clear detection of CO$_2$ at 4.3$\mu$m in an exoplanet, WASP-39b \citep{ahrer22}. The observations were made with the NIRSpec bright object time sequence mode and the 1.6” x 1.6” fixed slit aperture. A total of 21,500 integrations over 8.23 hours included the 2.8 hour transit duration.

\subsubsection{DD-ERS 1373: ERS observations of the Jovian System as a demonstration of JWST’s
capabilities for Solar System science}

JWST enables Solar System observations with moving-target tracking and several modes optimized for bright targets.
The pre-launch moving-target tracking requirement was 30 mas/sec, but the telescope managed to track the Double Asteroid Redirection Test (DART) impact on P/Didymos on 2022 September 26 at a rate of 105 mas/sec.
Program 1373 is an in-depth study of the Jovian system to characterize Jupiter’s cloud layers, winds, composition, auroral activity and temperature structure; to map Io and Ganymede, and to characterize Jupiter’s ring structure. The program uses all four instruments.

\subsubsection{DD-ERS 1386: High Contrast Imaging of Exoplanets and Exoplanetary Systems}

The direct characterization of exoplanets with JWST will enable mid-infrared coronagraphy and detailed spectroscopy for the first time. Program 1386 \citep{hinkley22} will use all four instruments to characterize two exoplanets and a circumstellar disk in the NIR and MIR. 
Early results include NIRCam and MIRI coronagraphic imaging of the super-Jupiter exoplanet HIP 65426b from $2\mu m$ to $16 \mu$m \citep{carter22}, the first direct detection of an exoplanet at wavelengths longer than $5\mu m$.  The observations are fit by a mass of 7.4$\pm$1.1 $M_{Jup}$. \cite{miles22} presented the highest fidelity spectrum to date of a planetary-mass object, VHS 1256 b, which is a $< 20 M_{Jup}$, widely separated, young brown dwarf companion. Water, methane, carbon monoxide, carbon dioxide, sodium and potassium were observed in the JWST spectra, indicating disequilibrium chemistry and clouds. They made a direct detection of silicate clouds for the first time in a planetary-mass companion.

\section{Mission Design}

The JWST mission consists of an observatory, a ground system provided by Space Telescope Science Institute and launch services provided by Arianespace under the direction of ESA. The observatory includes all the on-orbit hardware; the prime contractor for the observatory was Northrop Grumman Aerospace Systems (NGAS). The observatory consists of an optical telescope element provided by Ball Aerospace, a spacecraft and sunshield provided by NGAS and an Integrated Science Instrument Module (ISIM) constructed by Goddard Space Flight Center. The ISIM houses the four science instruments and provides them with thermal, electrical, structural, and data handling support.

The Near-Infrared Camera \citep[NIRCam;][this issue]{rieke23}, was built by Lockheed Martin under the direction of Principal Investigator Marcia Rieke of the University of Arizona. All of the near-infrared instruments include detectors from Teledyne Imaging Systems. The Near-Infrared Spectrograph \citep[NIRSpec;][this issue]{boker23} was built by EADS Astrium under the direction of ESA, and includes a microshutter assembly (MSA) and detector system built by Goddard Space Flight Center. The Near-Infrared Imager and Slitless Spectrograph \citep[NIRISS][this issue]{doyon23}, was built by Honeywell under the direction of CSA. The Fine Guidance System (FGS) is included with the NIRISS and shares the same optical bench. The Mid-Infrared Instrument \citep[MIRI;][this issue]{wright23} consists of an optical bench assembly built by a consortium of European countries organized by ESA, a cryocooler built by Northrop Grumman under the direction of Jet Propulsion Laboratories and a detector system by JPL and Raytheon Intelligence \& Space.

The mission design and its performance is summarized here. Details provided about the telescope are given by \cite{mcelwain23}, (this issue), construction, integration, and test of the observatory by \cite{menzel23}, (this issue), the on-orbit performance measured during commissioning by \cite{rigby23a}, (this issue), and the on-orbit backgrounds measured during commissioning by \cite{rigby23b}, (this issue).

\subsection{Launch, Orbit, Deployments, and Commissioning}

The JWST observatory was launched from {\it Centre Spatial Guyanais} at 12:20 UTC on 2021 December 25 by an Ariane 5 ECA+ rocket. The launch mass was $6161.4 kg$. The launch provided a near-perfect trajectory. Three mid-course correction (MCC) burns placed the observatory in an L2 halo orbit, approximately $1.5 \times 10^6 km$ from Earth. The successful launch and MCC burns used less on-board fuel than allocated; the remaining propellant will enable a fuel-limited lifetime of more than 20 years.

Launch was followed by more than 50 major deployments, which were completed successfully enroute to the final orbit. The major deployed systems included the solar panel, the high-gain antenna, the deployed tower assembly, the sunshield, the secondary mirror, an instrument radiator and the primary mirror wings. The sunshield deployment included 140 membrane release mechanisms, 70 hinge assemblies, 8 deployment motors, 400 pullies, and 90 cables totaling more than $400m$. Following the major deployments, which were completed in the first 14 days, the primary mirror segments and secondary mirror were moved off their launch locks. 

At the completion of the deployments, the telescope was pointed at HD 84406 \citep{gaia18} to begin the telescope alignment and phasing process. The major steps included (1) Segment Image Identification, (2) Segment Alignment, (3) Image Stacking, (4) Coarse Phasing, (5) Fine Phasing, (6) Telescope Alignment Over Instrument Fields of View, and (7) Iterate Alignment for Final Correction. Following telescope alignment, commissioning of the instruments included activating all instrument systems and commissioning the 17 science instrument modes. The final commissioning tasks included taking the Early Release Observations (ERO). Commissioning was completed by 2022 July 11 and 12 with the release of the EROs.

\subsection{Optical Telescope Element}

The optical telescope element (OTE) consists of a primary mirror made up of 18 hexagonal beryllium primary mirror segment assemblies (PMSAs), a 0.8m convex secondary mirror, and an aft optical assembly subsystem containing a tertiary mirror and a fine steering mirror. All the mirrors are coated with gold. The PMSAs and secondary mirror were aligned and phased on-orbit using actuators that had a total of 132 degrees of freedom. The total collecting area of the primary mirror is 25.4 $m^2$, as measured on-orbit using the NIRCam pupil imaging lens. The telescope was designed to be diffraction-limited at 2 $\mu m$ wavelength, defined as having a Strehl ratio $> 0.8$ \citep{bely03} at the end of a 5-year post-commissioning lifetime, equivalent to 150nm, root-mean squared (RMS) wavefront error (WFE). The WFE at the end of commissioning was $\sim 80$ nm RMS, equivalent to a diffraction limit at 1.1 $\mu m$. The expected degradation of the wavefront error due to micrometeoroid impacts and other effects over the operational lifetime of the mission will be closely monitored \citep[][this issue]{mcelwain23, rigby23a}.

\subsection{Spacecraft}

The spacecraft provides power, pointing, orbit maintenance, data storage and communications for the observatory. At the end of commissioning, the solar array provided an average of 1.5 kW of power with the ability to provide 3 kW when needed. The pointing system uses star trackers, inertial reference units (IRUs) containing gyros, reaction wheels, and a fine steering mirror (FSM) to point the telescope. There are three star trackers, one of which provides redundancy. There are two IRUs, one of which is redundant; each IRU includes four gyros, one of which provides redundancy. There are six reaction wheels, of which two provide redundancy. The star trackers, gyros and reaction wheels maintain the attitude and coarse pointing of the observatory. During fine guiding, after guide star acquisition, the FGS provides a 16 Hz positional update to the fine steering mirror to adjust the pointing of the telescope. At the end of commissioning, the pointing system delivers pointing stability of $\sim 1 mas$ (1$\sigma$ per axis), greatly exceeding the pre-launch estimates of $\sim 6 mas$. Orbit maintenance currently requires firing the on-board thrusters about once every 6 weeks. The thruster firings are also used to maintain angular momentum by de-spinning the reaction wheels. The frequency of orbit maintenance and angular momentum management is determined through ranging measurements and reaction wheel telemetry.

The on-board solid-state recorder holds 471 Gbits of science and engineering data. The data are downlinked via Ka band through the Deep Space Network (DSN) on a nominal schedule of two contacts per day totaling up to 12 hours. The actual downlink schedule varies from day to day with which antenna is in range and DSN scheduling with respect to other mission needs. Commands and observation plans are uplinked to the observatory through the DSN using S band.

\subsection{Sunshield and Cooling}

The sunshield consists of 5 layers of Kapton, about $14 m \times 22 m$ in size, that separate the $\sim 300K$ spacecraft from the telescope, and attenuate the $\sim$ 200 kW of incident Solar radiation to mW levels. The sunshield size and geometry provides an instantaneous field of regard of 40\% of the sky in an annulus that sweeps around the full sky once per year; each point on the sky is visible at least once in each six-month period. The telescope can point 5$\deg$ towards the Sun and 45$\deg$ away from the Sun, and can spin around the Sun-anti-Sun axis.

Immediately after launch, the observatory began passively cooling, a process that was completed within 120 days. The cooldown was controlled using heaters to ensure that the instruments would not be contaminated by condensation of outgassing water and other volatiles. The MIRI cryocooler (the only active cooling in the observatory) was turned on after the deployments and reached its final temperature on day 104. The final temperature reached by the secondary mirror is 29.2K. The primary mirror segments range from 34.7K to 54.5K, with the mirror segments closest to the core region near the sunshield at the bottom of the telescope warmer than the mirror segments at the top and wings. With these temperatures, JWST broad-band observations are background limited by zodiacal light out to about $12.5\mu m$ wavelength, and limited by thermal self-emission at longer wavelengths \citep[][this issue]{rigby23b}. The near-infrared instrument detector plane temperatures are actively maintained using heaters at 38.5K for NIRCam and NIRISS, and 42.8K for NIRSpec. The MIRI optical assembly is kept at 6K by the cryocooler, while the MIRI shield around the instrument is about 20K.

\section{Instruments}

JWST has four science instruments with a total of 17 science instrument modes (see Table~\ref{tab:instruments}). All of the instrument capabilities expected before launch have been enabled and are in use. Almost all of the instrument requirements have been exceeded; in particular most of the instrument modes are more sensitive than the pre-launch expectations, and the point spread function at the shorter wavelengths is sharper than the pre-launch expectations.

\begin{deluxetable*}{lcccc}
\tabletypesize{\small}
\tablecaption{Science Instrument Characteristics}
\tablewidth{\textwidth}
\tablehead{
\colhead{Instrument} &
\colhead{Wavelength ($\mu m$)} &
\colhead{Detector} &
\colhead{Plate Scale (mas/pix)} &
\colhead{Field of View}
}
\startdata
NIRCam&&&&\\
\hspace{0.25in}short&
0.6 - 2.3&
Eight 2048\ensuremath{\times}2048&
32&
2.2\ensuremath{\times}4.4 arcmin\\
\hspace{0.25in}long$^{\rm a}$&
2.4 - 5.0&
Two 2048\ensuremath{\times}2048
&
65&
2.2\ensuremath{\times}4.4 arcmin\\
&&&&\\
NIRSpec&
0.6 - 5.0&
Two 2048\ensuremath{\times}2048&
100&
\\
\hspace{0.25in}MSA$^{\rm b}$&
&
&
&
3.4\ensuremath{\times}3.1 arcmin\\
\hspace{0.25in}slits$^{\rm c}$&
&
&&
$\sim$200 mas $\times$ 4 arcsec\\
\hspace{0.25in}IFU&
&
&&
3.0$\times$3.0 arcsec\\
&&&\\
MIRI&&&&\\
\hspace{0.25in}imaging&
5 - 27&
1024\ensuremath{\times}1024&
110&
1.4\ensuremath{\times}1.9 arcmin\\
\hspace{0.25in}spectra$^{\rm d}$&
5 - 10&
&&26\ensuremath{\times}26 arcsec
\\
\hspace{0.25in}IFU&
5 - 28&
Two 1024\ensuremath{\times}1024&
200 to 470&
3.6\ensuremath{\times}3.6 
\\
&
&
&
&
to 7.5\ensuremath{\times}7.5 arcsec
\\
&&&\\
NIRISS&&2048\ensuremath{\times}2048&65&2.2\ensuremath{\times}2.2 arcmin\\
\hspace{0.25in}imaging&0.6 - 5.0&&&\\
\hspace{0.25in}WFSS&0.8 - 2.2&&&\\
\hspace{0.25in}SOSS&0.6 - 2.8&&&\\
\hspace{0.25in}AMI&2.8 - 4.8&&&\\
\\
FGS&0.6 - 5.0&Two 2048\ensuremath{\times}2048&65&2.2\ensuremath{\times}4.4 arcmin\\
&&&&
\enddata
\tablecomments{a) Use of a dichroic renders the NIRCam long-wavelength 
field of view co-spatial with the short wavelength channel, and 
the two channels acquire data simultaneously. (b) NIRSpec includes a 
micro-shutter assembly (MSA) with four 384$\times$175 micro-shutter arrays. The
individual shutters are each 250 (spectral) $\times$ 500 (spatial) mas.
(c) NIRSpec also includes several fixed slits which provide redundancy and 
high contrast spectroscopy on individual targets, and an integral field unit (IFU).
(d) MIRI includes a fixed slit for low-resolution (R$\sim$100) spectroscopy over the 5 to
10 $\mu$m range, and an integral field unit for R$\sim$3000 spectroscopy over the full 5
to 28 $\mu$m range. The long wavelength cut-off for MIRI spectroscopy is set by the
detector performance, which drops beyond 28.0 $\mu$m.}
\label{tab:instruments}
\end{deluxetable*}

\subsection{NIRCam}

NIRCam \citep[][this issue]{rieke03, horner04, rieke23} provides imaging from $0.6\mu m$ to $5.0\mu m$ in broad-band, medium-band and narrow-band filters. It has a wide-field slitless spectroscopy capability from $2.5\mu m$ to $5.0\mu m$. It has a coronagraphic mode. NIRCam is designed with two modules observing parallel fields of view; each module contains a dichroic at $2.4\mu m$ to provide simultaneous data in two filters longward and shortward of the dichroic. The two modules are identical, including the wavefront sensing hardware, and provide full redundancy in case of failure. The total field of view is $2.2 \times 4.4 arcmin^2$, and there are a total of ten detectors, with 8 in the short wavelength channel and 2 in the long wavelength channel. All of the near-infrared instruments (NIRCam, NIRSpec, NIRISS, and FGS) use H2RG HgCdTe $2048 \times 2048$ focal-plane arrays made by Teledyne Imaging Systems \citep[e.g.,][]{rauscher14}. The NIRCam plate scales are 32 milli-arsec per pixel in the short wavelength channels and 65 milli-arcsec per pixel in the long wavelength channels, Nyquist sampling the diffraction limit at $2.0\mu m$ and $4.0\mu m$, respectively. NIRCam also functions as part of the wavefront sensing and control system.

\subsubsection{NIRCam imaging}

NIRCam contains 2 extra-wide filters, 8 broad-band filters, 12 medium-band filters and 7 narrow-band filters. The broad-band filters span the full wavelength range of the instrument, the extra-wide and medium-band filters cover $1.0\mu m$ to $4.0\mu m$ and $1.4\mu m$ to $5.0\mu m$ respectively, and the narrow-band filters are selected to match individual spectral lines. NIRCam imaging sensitivity exceeds the pre-launch expectations in almost all filters. The requirements were 11.4nJy and 13.8nJy at $2.0\mu m$ and $3.5\mu m$, point-source sensitivity, 10$\sigma$ in 10,000s. The sensitivity at the end of commissioning were 7.3nJy and 8.8nJy respectively. NIRCam imaging is one of the most-used modes in Cycle 1 programs. An example program that uses this mode is Program 1963, a medium-band survey of the Hubble Ultra Deep Field.

\subsubsection{NIRCam wide-field slitless spectroscopy}

NIRCam wide-field slitless spectroscopy \citep[WFSS][]{greene17} provides $R \sim 1600$ spectra of all of the objects within the field of view using a grism. The grism is used in the long wavelength channel in combination with a wide or medium filter to provide spectroscopy in the $2.5\mu m$ to $5.0\mu m$ wavelength range. Short wavelength $< 2.5 \mu m$ imaging in the short wavelength channel can be taken simultaneously with the WFSS measurements. Commissioning data show that the total throughput is 20\% to 40\% higher than pre-launch expectations. An example using this mode is Program 2078, searching for galaxies at the same redshift as $6.5 < z < 6.8$ quasars. \citet{sun22a, sun22b} discovered H$\alpha$+[O III] $\lambda$5007 line emitters at $z>6$ using the NIRCam WFSS mode in commissioning data.

\subsubsection{NIRCam coronagraphy}

NIRCam coronagraphy \citep{krist09} has 3 round- and two bar-shaped coronagraphic masks for occulting a bright object. The inner working angles range from 0.40$\arcsec$ to 0.81$\arcsec$ for the round masks, corresponding to $6 \lambda / D$ at $2.1\mu m$, $3.35\mu m$ and $4.1\mu m$, and 0.13$\arcsec$ to 0.88$\arcsec$ for the bar masks. During a bar observation, the bright object is positioned behind the bar at the location where the IWA $\sim$ $4\lambda / D$. The masks are used in conjunction with a filter. Commissioning demonstrated that this mode provided a $5\sigma$ contrast at $1 \arcsec$ better than $4 \times 10^{-5}$ \citep{girard22}. An example using this mode is Program 1386, the DD-ERS high-contrast exoplanet imaging program \citep{hinkley22, carter22}. 

\subsubsection{NIRCam bright object time series - imaging}

Bright object time series (BOTS) observations are designed to measure photometric variations in relatively bright sources. NIRCam imaging BOTS uses rapid readout of sub-arrays ranging from 64$\times$64 to 160$\times$160, in combination with filters or a weak lens, to increase the readout cadence and increase the saturation limits. Commissioning observations showed that the NIRCam BOTS imaging performance was nominal. An example using this mode is Program 2635, studying infrared emission of 4U0142$+$61, a magnetar with a possible silicate spectral feature at 9.7$\mu m$, which has been interpreted as a passive disk surrounding the energetic isolated neutron star.

\subsubsection{NIRCam bright object time series - grism}

NIRCam grism BOTS observations provide $R \sim 1600$ spectroscopic observations of bright, isolated, time-varying sources. The spectroscopy in the long-wavelength channel is paired with weak lens observations in the short-wavelength channel to avoid saturation. This mode is capable of observing targets as bright as naked-eye stars (mag $<$ 5). Commissioning observations of the transiting exoplanet HAT-P-14 b obtained a 91 ppm spectrum (when binned to $R=100$). An example using this mode is Program 2084, searching for lava rain on the hot super-Earth planet 55 Cancri e.

\subsection{NIRSpec}

NIRSpec  \citep[][this issue]{jakobsen22, boker23} provides spectroscopy from 0.6 to 5.3 $\mu m$ at $R \sim 100$, $R \sim 1000$ and $R \sim 3000$ using fixed slits, a microshutter assembly (MSA) \citep{ferruit22}, or an integral field unit (IFU) \citep{boker22}. The detector system consists of two Teledyne 2048 × 2048 H2RG arrays controlled and read by SIDECAR ASICs. The $18 \mu m \times 18 \mu m$ pixels of the detector arrays project to an average of 0.103$\arcsec$ in the dispersion direction and 0.105$\arcsec$ in the spatial direction. Dispersion is done with a prism ($R = 30-300$) or gratings ($R = 500-1343$ or $R = 1321-3690$). The dispersion is crossed with filters to limit the bandwidth and resulting length of the spectra on the detectors. In most cases, the throughput of the instrument is higher than pre-launch expectations.

\subsubsection{NIRSpec multi-object spectroscopy}

NIRSpec multi-object spectroscopy is done using the MSA, which is a MEMS assembly consisting of four quadrants. Each MSA slit is 0.203$\arcsec$ by 0.463$\arcsec$, with a $\sim 0.07\arcsec$ wall between the openings. There are 730 (spectral) by 342 (spatial) pixels in the full array, spanning a field of view of approximately 3.6$\arcmin$ by 3.4$\arcmin$. The MSA is fully configurable, except for a limited number of failed slits. Typically, the MSA can be configured to observe up to 100 objects simultaneously, including sky subtraction, without overlapping spectra. Multi-object spectroscopy is one of the most highly used instrument modes in Cycle 1. An example using mode is Program 1345, The Cosmic Evolution Early Release Science (CEERS) Survey, the DD-ERS program targeting a deep field.

\subsubsection{NIRSpec fixed slit spectroscopy}

NIRSpec has five fixed slits, which provide the highest contrast and throughput on individual targets for NIRSpec. The fixed slits also provide a redundant spectroscopic capability to the mission if the MSA mechanism were to fail. Three fixed slits are $0.2 \arcsec$ wide by $3.3 \arcsec$ long, and one is $0.4 \arcsec$ wide by $3.8 \arcsec$ long. There is also a $1.6 \arcsec \times 1.6 \arcsec$ high-throughput slit that is primarily used with the NIRSpec bright object time series mode. An example using this mode is Program 1936, a target of opportunity program which will target a kilonova detected by the LIGO/Virgo/KAGRA gravitational wave detectors during their Observing Run 4.

\subsubsection{NIRSpec integral field unit spectroscopy}

The IFU entrance aperture is a contiguous $3.1 \arcsec \times 3.2 \arcsec$ field of view, divided into 30 slices totaling 900 spaxels, each $0.103 \arcsec \times 0.105 \arcsec$. By providing a full spectral data cube, the mode gives the most complete information on a single target in the near-infrared with JWST. The throughput is slightly lower than pre-launch expectations in the red, but higher in the blue. An example using this mode is 1355, Targeting Extremely Magnified Panchromatic Lensed Arcs
and Their Extended Star formation (TEMPLATES), the DD-ERS program that targets individual galaxies that are highly boosted by gravitational lensing.

\subsubsection{NIRSpec bright object time series}

NIRSpec BOTS \citep{birkmann22} primarily uses the $1.6 \arcsec \times 1.6 \arcsec$ fixed slit, combined with detector sub-arrays either 16 or 32 pixels wide, to rapidly monitor bright time-varying objects such as observations of stars with transiting exoplanets. The readout cadence can be as fast as 0.28s, potentially reaching stars brighter than $J<6$ in some modes. During commissioning, the mode was tested on HAT-P-14 b, and reached a noise level of $<$60 ppm \citep{espinoza22}. An example using this mode in Cycle 1 is Program 2159, following a hot super-Earth-size exoplanet for a full orbit to map the planet’s temperature.

\subsection{NIRISS}

NIRISS \citep[][this issue]{doyon12, doyon23} provides three specialized scientific capabilities and redundant broad-band imaging, over the wavelength range $0.7\mu m$ to $5.0 \mu m$. NIRISS is packaged with the FGS, which provides the signal to the fine steering mirror and the attitude control system to lock onto targets and provide fine guiding. NIRISS has a field of view of $2.2 \arcmin \times 2.2 \arcmin$, matching the FOV of one of the two NIRCam channels. NIRISS has a single 2048×2048 $5\mu m$ cut-off Hawaii-2RG detector, with 65 milliarcsec per pixel.

\subsubsection{NIRISS single object slitless spectroscopy}

NIRISS single-object slitless spectroscopy (SOSS) (Albert et al., in prep) defocuses the telescope beam to spread the signal from bright objects over about 25 pixels to avoid saturation. SOSS provides medium-resolution spectroscopy ($R \sim 70$) between 0.6$\mu m$ and 2.8$\mu m$. There are two usable orders. Using subarrays shortward of 1.0$\mu m$ allows targets as bright as $J = 6.5$ (Vega mag). An example is Program 2589, where the SOSS mode will be used to detect and characterize the possible atmospheres of the small, rocky exoplanets TRAPPIST 1b and 1c.

\subsubsection{NIRISS wide field slitless spectroscopy}

NIRISS wide-field slitless spectroscopy (WFSS) \citep{willott22} enables low-resolution ($R \sim 150$) slitless spectroscopy over the $2.2 \arcmin \times 2.2 \arcmin$ FOV at $0.8\mu m$ to $2.2 \mu m$ wavelength. It is optimized to search for Ly-$\alpha$ emitting galaxies during the epoch of reionization. It can also be used efficiently in parallel mode. Two orthogonal gratings provide dispersion in two directions to disentangle overlapping spectra and reduce confusion in crowded fields. The gratings are crossed with wide- or medium-band filters, which also reduces blending of the objects. Throughput of the WFSS mode exceeds the pre-launch expectations. An example is Program 1571, PASSAGE – Parallel Application of Slitless Spectroscopy to Analyze Galaxy Evolution, a pure-parallel search for active star-forming galaxies.

\subsubsection{NIRISS aperture masking interferometry}

NIRISS aperture masking interferometry \citep[AMI, ][]{sivaramakrishnan12, sivaramakrishnan22} uses a seven-aperture mask to enable high-contrast imaging at an inner working angles less than $\lambda/D$. AMI is used with the F380M, F430M, or F480M filters, and typically uses an $80 \times 80$ pixel subarray for bright sources. During commissioning, AMI was demonstrated by detecting AB Dor C, a companion separated by $\sim 0.3 \arcsec$ with a contrast ratio of 4.5 mag \citep{kammerer22}. An example is the DD-ERS Program 1349, WR DustERS, which will observe the Wolf-Rayet binary WR 137 with AMI to investigate the dust abundance, composition, and production rates of dusty sources in the colliding winds of the stars.

\subsubsection{NIRISS imaging}

NIRISS includes an imaging capability using a set of backup NIRCam broad-band filters and some medium-band filters. As NIRISS imaging covers half the FOV of NIRCam, and does not include a dichroic, this mode is primarily for imaging redundancy in the mission. NIRISS imaging can also be used in parallel to NIRCam imaging for additional areal coverage, and is used in support of WFSS data. Program 2561, which will observe the Frontier Field lensing cluster Abell 2744, uses NIRISS imaging in parallel to NIRCam imaging to increase the area of deep photometric studies of high-redshift galaxies at mild lensing magnifications.

\subsection{MIRI}

MIRI \citep[][this issue]{grieke15a, wright15, wright23} provides both imaging in broad-band filters and IFU spectroscopy from $5.0\mu m$ to $28.0\mu m$. It also has low-resolution slit spectroscopy from $5.0\mu m$ to $12.0\mu m$ (where the sensitivity is limited by the zodiacal light background) and a coronagraphic capability. MIRI has three arsenic-doped silicon (SI:As) impurity band conduction detector arrays, each of $1024 \times 1024$ pixel format with $25\mu m$ pixel pitch, made by Raytheon Intelligence \& Space \citep{ressler15, grieke15b}. The plate scale is 110 milli-arcsec per pixel, which Nyquist samples the point spread function at $6.25\mu m$. Two of the detectors are used for the medium-resolution spectroscopy, while the third is used for imaging, low-resolution spectroscopy and coronagraphy. The MIRI instrument is actively cooled to an operating temperature of $6.0K$ with a $\sim 6 K$/$18 K$ hybrid mechanical cooler, developed by Northrop Grumman in collaboration with JPL. The MIRI cooler uses gaseous helium as the coolant. There is a three stage Pulse-Tube Precooler which reaches $\sim 18 K$ and a fourth $\sim 6 K$ Joule-Thompson cooler stage. The cooler compressor is in the JWST spacecraft bus at room temperature, while the cold head assembly cooling the instrument is on the ISIM structure.

\subsubsection{MIRI imaging}

MIRI imaging \citep{bouchet15} uses 9 broad-band filters to cover the $5\mu m$ to $27\mu m$ wavelength region. The imaging field of view is $1.4 \times 1.9 arcmin^2$ sampled with $0.11\arcsec$ pixels. (The remaining field of view of the detector is occupied by the coronagraphs and the low-resolution spectrometer.) MIRI imaging can use sub-array readouts for bright objects that would saturate in a full frame, observing objects as bright as $0.1 Jy$ in the F560W filter \citep{glasse15}. An example using this mode is Program 2130, which will observe several square kpc in three nearby galaxies, M33, NGC 300 and NGC 7793, to measure dust-enshrouded stellar populations.

\subsubsection{MIRI low-resolution spectroscopy}

MIRI low-resolution spectroscopy \citep[LRS][]{kendrew15} provides $R \sim 100$ long-slit and slitless spectroscopy from $5\mu m$ to $12\mu m$, the MIR wavelength range where JWST observations are still zodiacal-light limited.
A slit mask is permanently in the field of view; slitless spectroscopy is available anywhere within the imager field of view when the $R\sim 100$ double prism assembly is selected in the imaging filter wheel. For bright sources, a sub-array readout can be used.
In practice, the source will be placed in a dedicated LRS slitless detector region and read out in a sub-array. It is expected that most LRS slitless targets will be bright nearby stars with transiting planets to obtain spectra of exoplanet atmospheres. Program 1658 will observe Pluto’s moon Charon using the MIRI LRS mode.

\subsubsection{MIRI medium-resolution spectroscopy}

MIRI medium-resolution spectroscopy \citep[MRS][]{wells15} provides integral-field spectroscopy over the full $5\mu m$ to $28\mu m$ MIRI wavelength range. The spectral resolution ranges from $\sim 3300$ at the short wavelength end to $\sim 1300$ at the longest wavelengths. There are four channels separated in wavelength by dichroics, with between 12 and 21 image slices. Depending on wavelength, the field of view ranges from $3.70\arcsec \times 3.70\arcsec$ to $7.74\arcsec \times 7.95\arcsec$. Each individual exposure provides two wavelength ranges on two detectors; three exposures are required to get a full wavelength spectrum. MIRI MRS is used in many programs; an example is Program 1549 which will observe three molecule-rich protoplanetary disks that were shown to have very bright water line emission in Spitzer spectra.

\subsubsection{MIRI coronagraphic imaging}

The imaging channel on MIRI includes four coronagraphs \citep{boccaletti15} for high-contrast imaging. The four coronagraphs are optimized for observations at $10.65\mu m$, $11.40\mu m$, $15.50\mu m$ and $\sim 23\mu m$. The short wavelength coronagraphs use four-quadrant phase masks \citep[4QPM;][]{rouan00, rouan07}, while the other is a more traditional Lyot design with an occulting spot in the image plane and a stop in the pupil plane. The 4QPMs are usable at a smaller inner working angle than more traditional designs, and can reach near $1\lambda/D$. Each of the 4QPMs provides a field of $24\arcsec \times 24\arcsec$, while the Lyot spot mask provides $30\arcsec \times 30\arcsec$ field of view. Each of the coronagraphs demonstrated a raw contrast ratio $>10,000$ at $6\lambda/D$ during commissioning. An example using this mode is Program 1618, which will search for planets and zodiacal dust around Alpha Centauri A.

\section{Science operations and proposal preparation}

JWST is controlled from a Mission Operations Center (MOC) at the Space Telescope Science Institute (STScI), which also runs the JWST science program. STScI issues annual calls for proposals (CFPs) for the General Observer programs. In between the CFPs, proposals for time-critical observations, or other observations that cannot be proposed to the annual call are considered for the Director’s Discretionary time. The scope of JWST’s competitively-selected programs range from large and Treasury programs that address multiple science goals and produce multi-use datasets to small programs that target important but specific science goals. All of the JWST data taken, including science programs and calibration data, are placed in the Mikulski Archive for Space Telescopes (MAST) at STScI and made available to the original proposers within a day or two of the data being taken. After an exclusive use period that ranges from 0 to 12 months depending on the type of program, the data are also freely available to other astronomers for archival research and other purposes.

In response to the annual CFP, proposals are prepared using the Astronomer’s Proposal Tool (APT) software package, which allows the proposer to specify both the textual proposal information (e.g., Title, Abstract, investigators, etc.) and the specifics of the observations. The APT is a sophisticated software package that ensures appropriate selection of observing parameters, checks the feasibility of the observations, and determines the times of the year that the observations could be scheduled, including planning guide star availability. APT also calculates the total allocated time needed for the observations. The text of the scientific justification and other proposal sections are attached to the proposal as a PDF within APT.

In addition to APT, observers will use the JWST Exposure Time Calculator (ETC) to determine many of the observation parameters, and to ensure that the observations reach the depth required for the science. The APT and ETC together contain sophisticated data simulation tools to visualize potential JWST observations. APT and the ETC are documented in an extensive series of on-line pages known as the JWST User Documentation \citep[JDox,][]{jdox}\footnote{For more information about proposing for JWST observing time or archival funding, see: \url{https://jwst-docs.stsci.edu/}}. JDox also documents the JWST data analysis tools.

\section{Getting JWST to space, what might we find, and what's next?}

Building on the inspiring and poetic 1996 HST and Beyond report of the Dressler committee, and with the vigorous support of NASA, ESA, and CSA leadership, the JWST team settled on the four top scientific priorities, documented the instrument and telescope performance requirements to meet scientific objectives, made plans, matured 10 technologies, made international agreements, and chose the instrument teams and contractors. The result is the world’s most powerful space telescope, performing better than expectations, with a projected lifetime of 20 years. We have reviewed the history, key technical choices, and we celebrate the people who made the observatory real.

We already see progress in the four key science themes. JWST has begun to address questions of the first galaxies and reionization by measuring spectroscopic redshifts of metal-poor galaxies beyond $z > 13$ \citep{curtis-lake22}, detecting multiple emission lines in a galaxy at $z = 10. 6$ \citep{bunker23}, and measuring galaxy luminosity functions at $z>7$ \citep{finkelstein22b}. JWST has studied galaxy assembly by detecting galaxy bars at $z>1$ \citep{guo23} and examining the quasar-galaxy connection at $z=2.94$ \citep{wylezalek22}. JWST has peered into star-forming regions to study the interactions between massive stars and the surrounding material \citep{reiter22} and measured the ice chemistry in a prestellar cloud \citep{mcclure23}. JWST has measured the temperature of a rocky exoplanet \citep{greene23}, and made the first detection of CO$_2$ in an exoplanet atmosphere \citep{ahrer22}. JWST observed the impact of NASA's Double Asteroid Redirection Test (DART) into asteroid Dimorphos and has made a detailed study of the Jovian system \citep{depater22}. Further JWST observations will continue to address the original science themes, and it is likely that the universe will surprise us with unexpected discoveries. Looking toward the future, our international teams have proven that extremely complex scientific space missions can be successful, paving the way towards the future great observatories recommended by the 2020 Decadal Survey.

\section{Acknowledgments}

The JWST mission is a joint project between the National Aeronautics and Space Agency, European Space Agency, and the Canadian Space Agency. The JWST mission development was led at NASA's Goddard Space Flight Center with a distributed team across Northrop Grumman Corporation, Ball Aerospace, L3Harris Technologies, the Space Telescope Science Institute, and hundreds of other companies and institutions. This mission was created by a team of people whose creativity and dedication made this scientific dream a reality.

Technical contributions were carried out at the Jet Propulsion Laboratory, California Institute of Technology, under a contract with the National Aeronautics and Space Administration (80NM0018D0004).

This work is based on observations made with the NASA/ESA/CSA James Webb Space Telescope. The data were obtained from the Mikulski Archive for Space Telescopes at the Space Telescope Science Institute, which is operated by the Association of Universities for Research in Astronomy, Inc., under NASA contract NAS 5-03127 for JWST.



\bibliographystyle{apj}
\bibliography{jwstreferences.bib}

\end{document}